\documentstyle[manuscript,prl,aps]{revtex}
\begin{document}
\tighten
\draft
\title{SUPERSONIC STRING MODEL FOR WITTEN VORTICES}
\author{Brandon Carter$^{2,3}$ and Patrick Peter$^{1,3}$}
\address{$^1$Department of Applied Mathematics and Theoretical Physics,\\
University of Cambridge, Silver Street, Cambridge CB3 9EW, England,\\
$^2$Isaac Newton Institute for Mathematical Sciences \\
20 Clarkson Road, Cambridge CB3 0EH, England,\\
$^3$D\'epartement d'Astrophysique Relativiste et de Cosmologie,\\
Observatoire de Paris-Meudon, UPR 176, CNRS, 92195 Meudon, France.}
\date{\today}
\preprint{\vbox{ \hbox{DAMTP-R94/56} \hbox{hep-ph/9411425}}}
\maketitle
\begin{abstract}
A new cosmic string model specified by two independent mass parameters
is introduced for the purpose of providing a realistic representation
of the macroscopic dynamical behaviour of Witten type
(superconducting) vortex defects of the vacuum. Unlike the self dual
single mass parameter models previously used for this purpose, the new
model successfully represents the effect of current saturation and the
feature that wiggle propagation remains supersonic even in the weak
current limit.
\end{abstract}
\pacs{PACS numbers: 98.80.Cq, 11.27+d}

The macroscopic dynamical behaviour of an ``ordinary" non conducting
Nielsen Olesen~\cite{1} vortex defect of the vacuum is well known to
be describable with sufficient precision for typical cosmological
applications~\cite{2} by a string model of the simple Goto-Nambu type.
However specific string models capable of providing an analogously
realistic description of the macroscopic dynamical behaviour of a
``superconducting" vacuum vortex of the kind originally introduced by
Witten~\cite{3} have so far been available only in a graphical or
tabulated numerical form~\cite{4}.

The general category~\cite{5} of elastic string models needed for this
purpose (among others) is expressible in terms of a 2-dimensional
worldsheet supported Lagrangian density that is expressible as the sum
of an electric coupling term, which in practice is commonly
unimportant and need not be considered here, and a {\it dominant
dynamic term} consisting of a ``master function" $\Lambda\{\chi\}$
depending only on the scalar $\chi=-h^{ab}\psi_{,a}\psi_{,b}$ where
$\psi_{,a}$ is the partial derivative with respect to 2-dimensional
coordinates $\sigma^a$ of an independently variable stream potential
function $\psi$ on the worldsheet, and $h_{ab}=g_{\mu\nu}x^{\mu}_{\,
,a} x^{\nu}_{\, ,b}$ is the surface metric induced by the
spacetime imbedding projection $\sigma^a \mapsto x^\mu$.  In any such
model there will also be a dual potential $\varphi$ say (proportional
to the phase of a complex scalar field in the underlying field theory)
whose gradient will be orthogonal to that of $\psi$, with magnitude
given by a scalar $\tilde\chi=-h^{ab}\varphi_{,a}\varphi_{,b}$ that is
dually related to $\chi$ as follows.

For any given master function there will be a corresponding ``dual"
master function $\tilde\Lambda\{\tilde\chi\}$ specified by
\begin{equation}\tilde\Lambda=\Lambda-{\cal K}\chi \hskip 1
cm\Leftrightarrow
\hskip 1 cm \Lambda=\tilde\Lambda-\tilde{\cal K}\tilde\chi\ ,
\label{1}\end{equation}
where
\begin{equation}{\cal K} =2{d\Lambda\over d\chi}=\tilde{\cal
K}^{-1}\hskip 1 cm
\Leftrightarrow
\hskip 1 cm \tilde{\cal K}=2{d\tilde\Lambda\over d\tilde\chi}={\cal
K}^{-1} \ ,\label{2}\end{equation} with \begin{equation}
\tilde\chi=-{\cal K}^2\chi\hskip 1 cm\Leftrightarrow
\hskip 1 cm \chi= -\tilde{\cal K}^2\tilde\chi\ . \label{3}\end{equation}
The field equations obtained by using $\Lambda$ as a Lagrangian with
$\psi$ and the embedding coordinate functions $x^\mu$ as the
independent variables will be the same~\cite{5} as those obtained by
using $\tilde\Lambda$ as Lagrangian with $\varphi$ and $x^\mu$ as the
independent variables.

Such a model generically provides not just one but {\it two distinct}
equations of state of simple elastic type relating the energy density
$U$ in the locally preferred rest frame of the string to the
corresponding string tension $T$, one equation applying to the range
where $\chi$ is positive and the other to the range in which $\chi$ is
negative. In the ``magnetic" range where $\chi$ is positive and
$\tilde\chi$ is negative, the stress energy momentum eigenvalues are
found to be given~\cite{5} by $U=-\tilde\Lambda$, and $T=-\Lambda$; in
this case there will be a corresponding conserved {\it number
density}, $\nu$, and an associated chemical potential or {\it
effective mass} variable, $\mu$, given by $\nu^2=-\tilde\chi$, and
$\mu^2=\chi$. On the other hand in the ``electric" range where $\chi$
is negative and $\tilde\chi$ is positive, the roles are interchanged:
one obtains $U=-\Lambda$, $T=-\tilde\Lambda$ and $\nu^2=-\chi$,
$\mu^2=\tilde\chi$. The foregoing construction ensures that, in each
range separately, the relations \begin{equation} U=T+\mu\nu\ ,\hskip
1cm
\mu={dU\over d\nu}\ ,\hskip 1 cm
\nu=-{dT\over d\mu} \ ,\label{4}\end{equation}
will automatically be satisfied. Moreover, in each of the two ranges,
the equation of state relation between $U$ and $T$ obtained in this
way will provide corresponding expressions~\cite{6}
\begin{equation}c_{_{\rm
E}}^2={T\over U}\ , \hskip 1 cm c_{_{\rm L}}^2 =-{dT\over
dU}={\nu\over\mu}{d\mu\over d\nu}\ , \label{5}\end{equation} for the
respective propagation speeds $c_{_{\rm E}} $ and $c_{_{\rm L}} $ of
{\it extrinsic} ``wiggle" perturbations of the worldsheet and {\it
sound type} (longitudinal) ``woggle" perturbations within the
worldsheet (in units such that the speed of light is unity).

The most serious early attempt at quantitative evaluation of the
functions $U$ and $T$ needed for the representation of Witten vortices
by string models of this type was made by Babul, Spergel, and
Piran~\cite{7} in the ``magnetic" range for particular values of the
relevant coupling constants in the underlying field theoretical model,
and a more extensive and numerically accurate analysis has since been
carried out by one of us~\cite{4}.  Prior to this work, most
investigations of the macroscopic dynamical behaviour were based on
the use~\cite{8,9,10} of a model given by
\begin{equation}\Lambda=-m^2+{\chi\over 2} \hskip 1 cm
\Leftrightarrow \hskip 1 cm
\tilde\Lambda=- m^2+{\tilde\chi\over 2} ,\label{6}\end{equation} where
$m$ is a constant having the dimensions of mass (of the order of the
Higgs mass scale in the underlying field theoretical model). This
naive ``folklore" model is the most obvious generalisation of the Goto
Nambu model whose action is simply constant,
$\Lambda=-m^2=\tilde\Lambda$. Like the trivial Goto Nambu model, the
naively linearised model (6) has the special property of being {\it
self dual} in the technical sense of providing an equation of state of
the {\it same} form, namely
\begin{equation}U+T=2m^2\ ,\label{7}\end{equation} both for positive
and negative values of $\chi$. Although some workers took this model
so seriously as to extrapolate it to the so called ``spring" limit of
vanishing tension $T$ (that is obtained for $\chi=2m^2$), it was
commonly understood that such a model could only be valid in the weak
current limit ($\vert\chi\vert\ll m^2$) in view of its obvious failure
to allow for the current saturation effect that had originally been
foreseen by Witten himself~\cite{3}. It has since been found to be
seriously misleading even for the weak current limit, because it can
be seen from (5) to be characterised by \begin{equation} c_{_{\rm
E}}<c_{_{\rm L}}=1\ ,\label{8}\end{equation} which means that it is
everywhere {\it subsonic} in the sense that extrinsic ``wiggles"
travel more slowly that longitudinal sound type ``woggles".  This
feature effectively disqualifies the ``folklore" model (6) even as an
approximate description of Witten type vortices, for which -- at least
in all the examples that have been investigated in detail so
far~\cite{4,7,11,12} -- the opposite has been found to be the case,
i.e.  the ``wiggles" actually travel {\it faster} than the ``woggles"
throughout the ``magnetic" range $\tilde\chi<0$ and for small and
moderate positive values of $\tilde\chi$ as well.

The discovery of supersonic wiggle propagation in Witten vortices has
potentially important cosmological implications in view of the
consideration first pointed out by Davis and Shellard~\cite{13} that
(unlike Goto Nambu models) conducting cosmic string loops have
stationary centrifugally supported equilibrium configurations which,
if they are sufficiently stable, could give rise to a catastrophic
cosmological mass excess, at least if the mass scale $m$ is that of
GUT symmetry breaking, though perhaps not~\cite{14} if it is only
that of electroweak symmetry breaking. The relevance to this of the
comparative speeds of ``wiggle" and ``woggle" propagation is that
stationary ring configurations of an elastic string loop have been
shown to be {\it alway stable}~\cite{15} with respect to classical
perturbations (which does not rule out the possibility of quantum
decay mechanisms whose treatment would be beyond the scope of the
macroscopic string description considered here) in any state of the
subsonic type, $c_{_{\rm E}}<c_{_{\rm L}}$ that always occurs for the
model (6).  However classical instability {\it can occur}~\cite{16,17}
for string loops of the supersonic kind that arises for Witten
vortices.

As an improvement on the model (6) that was used in most of the early
work on ``superconducting" cosmic strings, a minority of workers on
Witten vortices favoured an alternative model~\cite{18} characterised
by a worldsheet Lagrangian given by
\begin{equation}\Lambda=-m\sqrt{m^2-\chi}\hskip 1 cm\Leftrightarrow
\hskip 1 cm
\tilde\Lambda=-m\sqrt{m^2-\tilde\chi}\ ,\label{9}\end{equation}
that had been originally derived in a different physical context of a
rather artificial nature (on the basis of a Kaluza Klein type
projection mechanism) by Nielsen~\cite{19}. This model makes partial
allowance for the effect of current saturation, and as far as the weak
current limit was concerned it was commonly believed to be equivalent
to the more popularly favoured model (6). The naivity of that belief
is shown by the fact that while it shares with the more popular model
(6) the property of being self dual, in the sense of providing {\it
the same} equation of state for both positive and negative values of
$\chi$, the constant product form that is actually obtained~\cite{20}
from (9), namely \begin{equation}UT=m^4 \ ,\label{10}\end{equation}
can be seen to be characterised by
\begin{equation} c_{_{\rm E}}=c_{_{\rm L}} <1 \
,\label{11}\end{equation} i.e. the model (9) is of {\it permanently
transonic} type, in contrast with the more popular ``folklore" model
(6) which is of strictly subsonic type. The permanently transonic
character of the constant product equation of state (10) has been
shown to have extremely convenient mathematical consequences, leading
to explicit integrability by separation of variables for general (not
just equatorial) equilibrium states in a Kerr (and even Kerr-De
Sitter) rotating black hole background~\cite{21}.  Moreover in the
case of a flat background it leads to {\it complete} integrability not
just for stationary configurations but for general dynamically
variable string states~\cite{22}. As a by product of these desirable
properties, a string model of this permanently transonic type can be
used~\cite{22,23} to provide a smoothed average description of the
large scale behaviour of a simple Goto Nambu model (as characterised
by $T=U=m^2$) with a small scale wiggle structure whose details are
too complicated for it to be desirable or practical to provide a
complete fine grained description.  However contrary to what was too
hastily claimed~\cite{18}, this elegant transonic model can
emphatically {\it not} provide a qualitatively satisfactory
macroscopic description of supersonically wiggling Witten vortices.

The purpose of the present note is to present a new kind of analytic
model that is only slightly more complicated than the (self dual)
models described above, and that can provide a reasonably accurate
reproduction of the salient features of the numerically computed
equations of state derived for the specific examples of Witten
vortices that have been analysed so far~\cite{4,7,11,12}. Whereas the
self dual subsonic and transonic models (6) and (9) depend only on a
single constant mass mass parameter $m$, of the order of the Kibble
vacuum expectation value, the new kind of model introduced here
involves also a second such parameter $m_\ast\simeq\sqrt 3\, m_\sigma$
where $m_\sigma$ is the mass of the current carrier~\cite{3,4}. Its
most simple expression is given in terms of its dual master function
in the form \begin{equation}\tilde\Lambda=-m^2+{\tilde\chi\over
2}\Big(1-{\tilde\chi\over m_{\star}^2}
\Big)^{-1} \ .\label{12}\end{equation}
The range of validity of the model with this Lagrangian is limited
below by the condition that $c_{_{\rm L}}^2$ should be positive and
limited above by the condition that $c_{_{\rm E}}^2$ should be
positive, which means that $\tilde\chi$ must lie within the bounds
\begin{equation} -{1\over 3}<{\tilde\chi\over m_{\star}^2} <
1-{m_{\star}^2\over 2m^2+m_{\star}^2}\ .
\label{13}\end{equation} It can easily be verified that such a
model will be supersonic, with \begin{equation}c_{_{\rm L}}<c_{_{\rm
E}}<1\ ,\label{14} \end{equation} (except at $\tilde\chi=0$, where
$c_{_{\rm L}}=c_{_{\rm E}}=1$) throughout a range given by
\begin{equation}{\tilde\chi\over m_{\star}^2}<1-{3m_{\star}^2\over
2(2m^2+m_{\star}^2)}\ . \label{15}\end{equation} This range of
supersonicity will evidently include the weak current regime where
$\vert\tilde\chi\vert\ll m^2$ unless the second mass parameter exceeds
the bound \begin{equation} m_{\star} < 2m \ .\label{16}\end{equation}
The condition (16) will clearly be satisfied by a broad margin
whenever $m_{\star}^2 \ll  m^2$, as will be the case in the usual
examples~\cite{4} in which the carrier mass
$m_\sigma\simeq m_{\star}/\sqrt 3$ is assumed to be small compared with
the Kibble mass $\approx m$.

The upper bound in (13) arises immediately from the condition that the
tension must remain positive in the electric range where it is equal
to $-\tilde\Lambda$. Derivation of the lower bound in (13) requires
more detailed examination of the ``magnetic" range where
$\tilde\chi=-\nu^2\leq 0$, so that the Lagrangian (12) gives the
energy density $U$ and the effective mass $\mu$ (which is
proportional~\cite{5,20} to the current magnitude) in the form
\begin{equation} U=m^2+{\nu^2\over
2}\Big(1+{\nu^2\over m_{\star}^2}\Big)^{-1} \ ,
\hskip 1 cm  \mu=\nu\Big(1+{\nu^2\over m_{\star}^2}\Big)^{-2} \
.\label{17}\end{equation} The tension $T$ and longitudinal ``woggle"
speed $c_{_{\rm L}}$ are thus given by
\begin{equation}T=m^2-{m_{\star}^2\over 16}\big(1-c_{_{\rm L}}^4\big)\
, \hskip 1 cm c_{_{\rm
L}}^2=\Big(1-3{\nu^2\over m_{\star}^2}\Big)\Big(1+{\nu^2\over m_{\star}^2}
\Big)^{-1}\ .\label{18}\end{equation}
Hence the equation of state for $U$ as a direct function of $T$ is
\begin{equation}U=m^2+{m_{\star}^2\over 8}-{m_{\star}\over
2}\sqrt{{m_{\star}^2\over 16}-m^2+T}\ ,
\label{19}\end{equation}
in which the choice of sign in the root is dictated by the condition
$\nu^2 < m_{\star}^2/3$ expressing the local stability requirement
that the ``woggle" speed given by (18) should be real, which is what
fixes the lower bound in (13). It can be seen that the allowed range
for the tension in the ``magnetic'' range will satisfy the ``no
spring" condition~\cite{12} of having a lower bound (namely
$T=m^2-m_{\star}^2/16$) that is {\it strictly} positive provided
$m_{\star}<4m$, which will evidently hold in any model satisfying the
stronger condition (16) that one would generally expect to be
satisfied by a broad margin in practice.

Since the model (12) does not share with (6) and (9) the property of
self duality, (17), (18) and (19) do not apply directly to the
``electric" range, but the corresponding results are obtainable by an
entirely analogous application of the principles recapitulated above.
In particular, instead of (19), the equation of state giving the
energy density as a function of tension in the ``electric" range
$\tilde\chi>0$ is found to have the simple quadratic form
\begin{equation}U=2m^2\Big(1+2{m^2\over
m_{\star}^2}\Big)-\Big(1+8{m^2\over m_{\star}^2}\Big)T +{4\over
m_{\star}^2} T^2 \ .\label{20}\end{equation}

It is to be remarked in conclusion that the qualitative behaviour (and
in particular the form of the dynamical equations of motion) of the
new string model will be fully determined just by the single
dimensionless parameter ratio $m_{\star}^2/m^2$, for which the
appropriate value is expected to be very small, $m_{\star}^2/m^2\simeq
3 m_\sigma^2/m^2\ll 1$. The unsuitability of the undeservedly popular
``folklore" model (6) can be seen from the fact that the corresponding
linear equation of state (7) is obtained from the mutually dual
equations of state (20) and (21) of the new model in the {\it
opposite} limit, $m_{\star}^2/m^2\rightarrow\infty$.

The authors wish to thank Xavier Martin for discussions.

\end{document}